\documentclass[11pt]{article}
\usepackage{amsmath,amsfonts,amsthm,amssymb}
\usepackage[hmargin={1.0in, 1.0in}, vmargin={1.0in, 1.0in}]{geometry}
\usepackage{setspace}
\usepackage[utf8]{inputenc}
\usepackage[english]{babel}
\usepackage{fancyhdr}
\usepackage{lastpage}
\usepackage{extramarks}
\usepackage{chngpage}
\usepackage{soul,color}
\usepackage{graphicx,float,wrapfig}
\usepackage{layout}
\usepackage{fullpage}
\usepackage{natbib}
\usepackage{accents}
\usepackage[titletoc]{appendix}
\usepackage{xr}
\usepackage{longtable}
\usepackage{subcaption}
\usepackage{array}
\usepackage{algorithm}
\usepackage{titlesec}
\titlespacing*\section{0pt}{8pt plus 2pt minus 2pt}{0pt plus 2pt minus 2pt}
\titlespacing*\subsection{0pt}{6pt plus 2pt minus 2pt}{0pt plus 2pt minus 2pt}
\titlespacing*\subsubsection{0pt}{2pt plus 2pt minus 2pt}{0pt plus 2pt minus 2pt}

\newcolumntype{H}{>{\setbox0=\hbox\bgroup}c<{\egroup}@{}}
\usepackage[noend]{algpseudocode}

\makeatletter
\def\BState{\State\hskip-\ALG@thistlm}
\makeatother

\def\mC{\mathcal C}

\def\mT{\mathcal T}
\def\iP{\textit{P}}

\def\bftau{\mathbb{\pmb{\tau}}}

\DeclareMathOperator*{\argmin}{arg\,min}

\newtheorem{theorem}{Theorem}
\newtheorem{assumption}{Assumption}

\newtheorem{proposition}{Proposition}

\newtheorem*{definition*}{Definition}
\theoremstyle{definition}

\newcommand{\blind}{1}

\begin{document}
\setlength{\abovedisplayskip}{4pt}
\setlength{\belowdisplayskip}{4pt}
\setlength{\abovedisplayshortskip}{2pt}
\setlength{\belowdisplayshortskip}{2pt}

\if1\blind
{
	\title{Alternating Pruned Dynamic Programming for Multiple Epidemic Change-Point Estimation}	
	\author{
		Zifeng Zhao\\
		Mendoza College of Business \\ 
		University of Notre Dame\\
		\and
		Chun Yip Yau\\
		Department of Statistics\\
		Chinese University of Hong Kong\\
	}
	\date{}	
	\maketitle
} \fi

\if0\blind
{
	\title{Alternating Pruned Dynamic Programming for Multiple Epidemic Change-Point Estimation}
	\author{}
	\date{}
	\maketitle
	\vspace{-2cm}
} \fi

\begin{abstract}
	In this paper, we study the problem of multiple change-point detection for a univariate sequence under the epidemic setting, where the behavior of the sequence alternates between a common normal state and different epidemic states. This is a non-trivial generalization of the classical (single) epidemic change-point testing problem. To explicitly incorporate the alternating structure of the problem, we propose a novel model selection based approach for simultaneous inference on both change-points and alternating states. Using the same spirit as profile likelihood, we develop a two-stage alternating pruned dynamic programming algorithm, which conducts efficient and exact optimization of the model selection criteria and has $O(n^2)$ as the worst case computational cost. As demonstrated by extensive numerical experiments, compared to classical general-purpose multiple change-point detection procedures, the proposed method improves accuracy for both change-point estimation and model parameter estimation. We further show promising applications of the proposed algorithm to multiple testing with locally clustered signals, and demonstrate its advantages over existing methods in large scale multiple testing, in DNA copy number variation detection, and in oceanographic study.
\end{abstract}

\noindent
{\it Keywords:} Change-point, DNA copy number variation, Epidemic alternative, Model selection, Multiple testing

\section{Introduction}
Change-point detection is an active research area in statistics and has been studied extensively due to its broad applications in many fields such as bioinformatics, climate science, finance, and genetics among others. There is vast literature in change-point detection, see \cite{Yao1987}, \cite{Davis2006}, \cite{Aue2009}, \cite{Killick2012}, \cite{Chan2013}, \cite{Zou2014}, \cite{Matteson2014}, \cite{Yau2016}, \cite{Fearnhead2017}, \cite{Jiang2020} and references therein.

A less studied yet important type of change-point detection problem is the epidemic change-point detection, which is first proposed and studied in \cite{Levin1985}. Let $y_{1:n}=(y_1,\ldots,y_n)$ be a sequence of independently distributed univariate observations. Roughly speaking, under the (classical) epidemic change-point setting, there exist two change-points $0<\tau_1<\tau_2<n$ such that $y_{1:\tau_1}$ and $y_{(\tau_2+1):n}$ follow the same distribution and $y_{(\tau_1+1):\tau_2}$ follows a different distribution. The two segments on the sides $y_{1:\tau_1}$ and $y_{(\tau_2+1):n}$ are referred to as the normal state and the central segment $y_{(\tau_1+1):\tau_2}$ is referred to as the epidemic state. We call this setting the (classical) single epidemic change-point setting.

In the literature, the epidemic change-point detection is typically formulated as a hypothesis testing problem, where different test statistics have been proposed to test the null hypothesis of no change-point against the above defined epidemic alternative with two change-points, see \cite{Yao1993}, \cite{Guan2004}, \cite{Arias-Castro2005} and \cite{Aston2012}. Moreover, the existing literature focus on the single epidemic change-point setting where the data is assumed to start at the normal state and only one \textit{single} epidemic state is allowed. The more realistic setting of detecting multiple epidemic change-points, however, is largely unexplored. One exception is a recent work by \cite{Fisch2019}, where the authors consider the context of multiple epidemic changes but with a focus on anomaly and outliers detection.

In this paper, we propose a model selection based framework on multiple epidemic change-points estimation. Specifically, we assume that under the epidemic alternative, there exist $m$ unknown change-points $0<\tau_1<\tau_2<\ldots<\tau_m<n$ such that the distribution of $y_t$ alternates between a (common) normal state and (different) epidemic states. For a concrete example, let the number of change-points $m$ be even and the data $y_{1:n}$ start at the normal state. Denote $\tau_0=0$ and $\tau_{m+1}=n$. Under the multiple epidemic change-point setting, the $m/2+1$ odd-numbered segments $y_{(\tau_{2k}+1) : \tau_{2k+1}}, k=0,\ldots, m/2$ are at the (common) normal state and the $m/2$ even-numbered segments $y_{(\tau_{2k-1}+1) : \tau_{2k}}, k=1,\ldots, m/2$ are at (different) epidemic states. In other words, the data $y_{1:n}$ alternates between the normal state and epidemic states.

The multiple epidemic change-point setting incorporates the aforementioned single epidemic setting as a special case and is more realistic in that it allows the observations $y_{1:n}$ to move back and forth between the common normal state and epidemic states. One motivating example for multiple epidemic change-point setting is {the DNA copy number variation~\citep[see][]{Olshen2004,Niu2012,Xiao2015,Shin2020}}, where the observations $y_{1:n}$ are the log-ratios of the copy number of genes between the test and reference sequence. For most genes, there is no variation~(common normal state) and the mean log-ratio is a common constant~(e.g. 0). When there is variation~(epidemic state), depending on the duplication or deletion of certain genes, the mean log-ratio can be either larger or smaller than that of the normal state. Another important example is large scale multiple testing with locally clustered signal as considered in \cite{Cao2015}, where a sequence of $p$-values $p_{1:n}$ are observed with $p_i$ being the $p$-value for the $i$th test, and we need to perform $n$ hypothesis tests based on $p_{1:n}$. The signal is locally clustered in the sense that the sequence of $p$-values can be partitioned into alternating blocks of signal~(epidemic state, where $p$-values do not follow $U(0,1)$) and noise~(common normal state, where $p$-values follow $U(0,1)$). The two examples are later discussed in detail in Section 5.

Compared to the conventional multiple change-point detection problem, the unique aspect of the multiple epidemic change-point setting is that there is an underlying alternating structure on the behavior of the observation $y_{1:n}$. Same as standard change-point problems, our primary interest is to recover the unknown number and locations of change-points. {Distinctive from standard change-point problems}, a further interest is to recover the underlying alternating states of $y_{1:n}$. Specifically, the task is to assign a normal or epidemic label to each estimated segment.

The unique alternating structure of states and the shared common behavior among all normal state segments impose both challenges and opportunities for change-point detection. Specifically, existing {general-purpose} multiple change-point detection algorithms such as CBS in \cite{Olshen2004}, PELT in \cite{Killick2012} and WBS in \cite{FRYZLEWICZ2014} cannot directly recover the underlying alternating states and thus require additional post-analysis on the estimation result. On the other hand, intuitively, if an algorithm can explicitly incorporate and exploit the alternating structure and the knowledge that segments at the normal state share the same behavior, improved estimation accuracy should be expected due to the additional information on the constrained structure of the problem. {As an illustrative example, Proposition S.1 of the supplementary material shows that incorporating the structure of epidemic change can help uniformly improve the power of a change-point test.}

Motivated by the above observations, in this paper, we propose a novel alternating dynamic programming algorithm, named aPELT, to efficiently solve the multiple epidemic change-point problem. The proposed approach is based on the seminal work of PELT in \cite{Killick2012}, but involves an explicit treatment for the alternating structure and common normal state behavior.

The advantages of aPELT are three-fold. First, by incorporating the shape-constraint explicitly, aPELT achieves \textit{simultaneous} inference on both change-points and alternating states of the sequence, thus does not require any post processing of the estimation result. Second, as demonstrated by extensive numerical experiments and real data applications, the explicit treatment for alternating states further helps to improve accuracy for both change-point estimation and parameter estimation. The proposed aPELT has meaningful applications in multiple testing problems with locally clustered signals and in DNA copy number variation detection, where favorable performance over existing general-purpose methods is observed~(see Section 5 for more details). {Third, similar to PELT, it can be applied to detect {various types of changes (such as mean, variance)} under a range of statistical criteria such as likelihood, quasi-likelihood and cumulative sum of squares, and further enjoys computational efficiency, thus aPELT can be used to segment large datasets.}

A related yet rather different stream of literature is the constrained dynamic optimization in \cite{Hocking2015} and \cite{Hocking2018}. Motivated by mean changes in ChIP-seq data, the authors propose efficient algorithms~({PeakSeg and GFPOP}) to solve a model selection problem under the constraint that a decrease in mean must be followed by an increase, and vice versa. In contrast, our multiple epidemic change-point setting does not impose any directional relation on the normal state and epidemic state behavior, and thus cannot be covered by the setting in \cite{Hocking2015} and \cite{Hocking2018}.\footnote{For example, consider a sequence that has the following alternating state changes: a high (epidemic) state $\to$ the normal state $\to$ a low (epidemic) state $\to$ the normal state $\to$ another high (epidemic) state.} Moreover, our use of common normal state parameter poses further difficulty on the optimization, since the constraint is not only on two neighboring segments but instead on all normal state segments.

The rest of the paper is organized as follows. In Section 2, we formulate the multiple epidemic change-point detection problem and review the model selection approach for general change-point problems. For a tailored and efficient solution, an alternating dynamic programming algorithm~(aPELT) is proposed in Section 3. The efficiency and accuracy of the proposed method are demonstrated via extensive numerical experiments in Section 4. Applications of aPELT to DNA copy number variation and multiple testing with locally clustered signals are presented in Section 5. The paper concludes with a discussion. Additional simulation, real data application and technical materials can be found in the supplementary material.

\section{Background and Existing Solutions}
\subsection{Basic setting}
Roughly speaking, change-point detection can be considered as the identification of points within a dataset where the statistical properties change. In this paper, we assume that $y_{1:n}=(y_1,\ldots,y_n)$ is a sequence of independently distributed univariate observations. There are $m$ change-points $0<\tau_1\ldots<\tau_m<n$ that split the data into $m+1$ segments. Define $\tau_0=0$ and $\tau_{m+1}=n$, we have that the $i$th segment contains data $y_{(\tau_{i-1}+1):\tau_i}$.

We assume that the distribution of $y_t$ belongs to a parametric family $f(y|\theta, \gamma)$, where $\theta\in \mathbb{R}^{p_1}~(p_1\geq 1)$ denotes the parameter of interest and $\gamma\in \mathbb{R}^{p_2}~(p_2\geq 0)$ denotes the nuisance parameter. For example, in mean change detection for independent Gaussian observations, $\theta$ is the mean and $\gamma$ is the variance of $y_t$, and in change-point detection for independent Poisson counts, $\theta$ is the intensity of $y_t$ and there is no $\gamma.$ Denote the parameters for the $i$th segment $y_{(\tau_{i-1}+1):\tau_i}$ as $(\theta^i, \gamma^i)$, we have $\theta^i\neq \theta^{i+1}$ for $i=1,\ldots,m.$ Note that we do not put any restriction on the nuisance parameters $\gamma^i$ except assuming that they are unknown.

Under the multiple epidemic change-point setting, we further assume that $\{\theta^i\}_{i=1}^{m+1}$ alternates between a common normal state $\theta^o$ and different epidemic states. More formally, for any $i=1,\ldots,m$, if the $i$th segment $y_{(\tau_{i-1}+1):\tau_i}$ follows $f(y|\theta^o)$~(or $f(y|\theta)$ for some $\theta\neq \theta^o$), then the $(i+1)$th segment $y_{(\tau_{i}+1):\tau_{i+1}}$ follows $f(y|\theta)$ for some $\theta\neq \theta^o$~(or $f(y|\theta^o)$). The only requirement for an epidemic state is that $\theta \neq \theta^o$ without any directional constraint. For the multiple epidemic change-point setting, our inference interests are two-fold: 1. to recover the unknown number and locations of change-points, 2. to recover the alternating states of the observation $y_{1:n}$.

As mentioned in Section 1, existing literature focus on single epidemic change-point detection with $m=2$ and assume data starts at the normal state. Under such setting, typically a test statistic of the form $\max_{1\leq i<j<n}Z(i,j)$ is constructed for change-point estimation via hypothesis testing. With $m$ and initial state of the data being unknown, a direct generalization of this testing procedure to the multiple epidemic change-point setting is not obvious, {due to the constraint of alternating states and common normal state behavior}. Thus, we instead tackle the multiple epidemic change-point detection via a model selection approach.

\subsection{Optimal Partitioning and PELT}
The multiple epidemic change-point detection is a special type of multiple change-point detection problem. In this section, we review two existing model selection based detection algorithms, which serves as the basis for our proposed alternating change-point detection procedure. For the moment, assume that we are doing classical multiple change-point detection, thus the only requirement is $\theta^i\neq \theta^{i+1}$ for $i=1,\ldots,m$.

Given the observation $y_{1:n}$, denote $\mT(n)=\{\bftau: 0=\tau_0<\tau_1,\ldots<\tau_{m}<\tau_{m+1}=n\}$ as the candidate set of all possible vectors of change-points. The model selection approach estimates the true change-points $\bftau^o$ by minimizing a penalized loss function 
\begin{align}
F(n)=\min_{\bftau \in \mT(n)}\left\{\sum_{i=1}^{m+1}[\min_{\theta^i,\gamma^i}\mC(y_{(\tau_{i-1}+1):\tau_i}|\theta^i,\gamma^i)+\iP]\right\}=
\min_{\bftau \in \mT(n)}\left\{\sum_{i=1}^{m+1}[\mC(y_{(\tau_{i-1}+1):\tau_i})+\iP]\right\},
\label{MDL}
\end{align}
where $\mC(y_{(\tau_{i-1}+1):\tau_i}|\theta^i,\gamma^i)=\sum_{t=\tau_{i-1}+1}^{\tau_i}g(y_t|\theta^i,\gamma^i)$ denotes the measure of model fit such as twice the negative log-likelihood, $\mC(y_{(\tau_{i-1}+1):\tau_i})=\min_{\theta^i,\gamma^i}\mC(y_{(\tau_{i-1}+1):\tau_i}|\theta^i,\gamma^i)$, and $\iP$ denotes the penalty for model complexity such as BIC or MDL, see \cite{Aue2011} and \cite{Aue2014}.

The optimization of \eqref{MDL} is in general difficult {due to the $L_0$ penalty}. Using dynamic programming, \cite{Jackson2005} propose the Optimal Partitioning~(OP) algorithm which obtains the exact solution of \eqref{MDL} with $O(n^2)$ computational complexity. The essential idea of OP is the recursive relationship where for any $s\leq n$,
\begin{align}
F(s)=&\min_{\bftau \in \mT(s)}\left\{\sum_{i=1}^{m+1}[\mC(y_{(\tau_{i-1}+1):\tau_i})+\iP]\right\}\nonumber\\
=&\min_{t<s}\left\{\min_{\bftau \in \mT(t)}\sum_{i=1}^{m}[\mC(y_{(\tau_{i-1}+1):\tau_i})+\iP]+\mC(y_{(t+1):s})+P\right\}\label{recursive}\\
=&\min_{t<s}\left\{F(t)+\mC(y_{(t+1):s})+P\right\}\nonumber.
\end{align}
This provides a recursion which gives the minimum cost $F(s)$ of $y_{1:s}$ in terms of the minimum cost $F(t)$ of $y_{1:t}$ for $t<s$, and thus $F(n)$ can be solved in turn for $s=1,2,\ldots,n.$ Note that the essential condition for the recursion \eqref{recursive} to hold is that the optimization of $\mC(y_{(t+1):s})=\min_{\theta,\gamma}\sum_{i=t+1}^{s}g(y_i|\theta,\gamma)$ is independent across different segments~(Bellman's principle of optimality), which is true under the classical multiple change-point setting.

Assuming the existence of a constant $K$ such that for all $t<s<n$, $\mC(y_{(t+1):s})+\mC(y_{(s+1):n})+K\leq \mC(y_{(t+1):n})$, \cite{Killick2012} propose the PELT algorithm, which further reduces the computational complexity of OP and can solve $F(n)$ in linear time under {certain conditions}. The key idea of PELT is that for the calculation of $F(s)$, we do not need to consider all $\{t|t<s\}$ but only a pruned subset $\big\{t|t<s\big\} \mathbin{\big\backslash} \big\{t|\text{there exists } t<t'<s \text{ such that } F(t)+\mC(y_{(t+1):t'})+K\geq F(t') \big\}$, and thus achieve a lower computational cost.

\section{Multiple Epidemic Change-point Detection via Alternating PELT}
Compared to the classical setting, the multiple epidemic change-point setting imposes an implicit shape-constraint on the model parameter $\{\theta^i\}_{i=1}^{m+1}$ such that $\{\theta^i\}_{i=1}^{m+1}$ alternates between the normal state and epidemic states, and all normal state $\theta^i$s are the same.

{As mentioned in the introduction, besides segmenting the sequence via change-point detection, another primary interest is to simultaneously recover the label of each segment.} In other words, we would like the estimated parameter $\{\hat{\theta}^i\}_{i=1}^{m+1}$ to possess the alternating structure. However, neither OP nor PELT can directly recover the alternating structure since the shape-constraint is not explicitly considered in the penalized loss function $F(n)$ in \eqref{MDL}, where $\bftau$ only determines the number and locations of the change-points but does not restrict the state of each segment.

To impose the alternating structure of $\{\theta^i\}_{i=1}^{m+1}$, we propose to modify the penalized loss function in \eqref{MDL} by explicitly assigning states to segments. Note that due to the alternating structure, for any given $\bftau$, once the state of the last segment is determined, the states of all other segments are fixed accordingly. Thus, given $m$ and $\bftau$, we define four index sets
\begin{align*}
&\mathcal{S}^{oo}_{m}=\{i|i\text{th segment normal}, (m+1)\text{th segment normal}\},\\
&\mathcal{S}^{o1}_{m}=\{i|i\text{th segment epidemic}, (m+1)\text{th segment normal}\},\\
&\mathcal{S}^{1o}_{m}=\{i|i\text{th segment normal}, (m+1)\text{th segment epidemic}\},\\
&\mathcal{S}^{11}_{m}=\{i|i\text{th segment epidemic}, (m+1)\text{th segment epidemic}\}.
\end{align*}
Depending on the state of the last segment, the four index sets assign states and group the segments of $\bftau$ by normal or epidemic states. For example, for $m=5$, $\mathcal{S}^{oo}_{5}=\{2,4,6\}$, $\mathcal{S}^{o1}_{5}=\{1,3,5\}$, $\mathcal{S}^{1o}_{5}=\{1,3,5\}$ and $\mathcal{S}^{11}_{5}=\{2,4,6\}$.

Based on the four index sets, we further define two penalized loss functions
\vspace{-1cm}
\begin{changemargin}{-0.3cm}{-0.5cm}
\begin{align}
&F^*_o(n)=\min_{\bftau \in \mT(n)}\left\{\min_{\theta^o}\sum_{i \in\mathcal{S}^{oo}_m} [\min_{\gamma^i}\mC(y_{(\tau_{i-1}+1):\tau_i}|\theta^o,\gamma^i)+\iP^o] + \sum_{i \in\mathcal{S}^{o1}_m} [\min_{\theta^i,\gamma^i}\mC(y_{(\tau_{i-1}+1):\tau_i}|\theta^i,\gamma^i)+\iP^1]\right\},\label{MDL1}\\
&F^*_1(n)=\min_{\bftau \in \mT(n)}\left\{\min_{\theta^o}\sum_{i \in\mathcal{S}^{1o}_m} [\min_{\gamma^i}\mC(y_{(\tau_{i-1}+1):\tau_i}|\theta^o,\gamma^i)+\iP^o] + \sum_{i \in\mathcal{S}^{11}_m} [\min_{\theta^i,\gamma^i}\mC(y_{(\tau_{i-1}+1):\tau_i}|\theta^i,\gamma^i)+\iP^1]\right\},\label{MDL2}
\end{align}
\end{changemargin}
where $\iP^o$ denotes the penalty for the segment at normal state and $\iP^1$ denotes the penalty for the segment at epidemic state. {Note that $\iP^o$ and $\iP^1$ typically take different values, as the penalty for $\theta^o$ should be lower due to the fact that it is common across all normal state segments.}

By design, $F^*_o(n)$ forces the last segment of $\bftau$ to be at the normal state and $F^*_1(n)$ forces the last segment to be at the epidemic state. Moreover, $F^*_o(n)$ and $F^*_1(n)$ explicitly incorporate the alternating structure of $\{\theta^i\}_{i=1}^{m+1}$ since $\mathcal{S}_m$ enforces alternating normal and epidemic states among segments and all segments in $\mathcal{S}^{oo}_m$ or $\mathcal{S}^{1o}_m$ share a common $\theta^o$. Thus, for the simultaneous inference of change-points and alternating states, we can then solve the modified penalized loss function
\begin{align}
F^*(n)=\min(F^*_o(n),F^*_1(n)),
\label{MDLnew}
\end{align}
which explicitly incorporates the alternating shape-constraint of $\{\theta^i\}_{i=1}^{m+1}$ and requires no prior knowledge of the initial state of $y_{1:n}$.

To solve \eqref{MDLnew} efficiently, a recursive relationship similar to \eqref{recursive} is needed for a dynamic programming based algorithm. However, due to the presence of the common parameter $\theta^o$ across all normal state segments, the recursive relationship in \eqref{recursive} no longer holds since the optimization of $\sum_{i \in\mathcal{S}^{oo}_m} [\min_{\gamma^i}\mC(y_{(\tau_{i-1}+1):\tau_i}|\theta^o,\gamma^i)+\iP^o]$ in $F^*_o(n)$ and $\sum_{i \in\mathcal{S}^{1o}_m} [\min_{\gamma^i}\mC(y_{(\tau_{i-1}+1):\tau_i}|\theta^o,\gamma^i)+\iP^o]$ in $F^*_1(n)$ are no longer independent across segments. Thus, the previous algorithms break down and a new algorithm is needed for the computationally feasible estimation.

\subsection{A two stage optimization procedure}
The key observation is that the optimization of the common $\theta^o$ causes the breakdown of the recursion \eqref{recursive}. To bypass this obstacle, we propose a two-stage optimization procedure for $F^*(n)$, which separates the optimization of $\theta^o$ and other model parameters $(\gamma, \bftau)$. This procedure shares the same spirit as profile likelihood. For any fixed $\theta^o$, we define
\begin{align*}
&F^*_o(n;\theta^o)=\min_{\bftau \in \mT(n)}\left\{\sum_{i \in\mathcal{S}^{oo}_m} [\min_{\gamma^i}\mC(y_{(\tau_{i-1}+1):\tau_i}|\theta^o,\gamma^i)+\iP^o] + \sum_{i \in\mathcal{S}^{o1}_m} [\min_{\theta^i,\gamma^i}\mC(y_{(\tau_{i-1}+1):\tau_i}|\theta^i,\gamma^i)+\iP^1]\right\},\\
&F^*_1(n;\theta^o)=\min_{\bftau \in \mT(n)}\left\{\sum_{i \in\mathcal{S}^{1o}_m} [\min_{\gamma^i}\mC(y_{(\tau_{i-1}+1):\tau_i}|\theta^o,\gamma^i)+\iP^o] + \sum_{i \in\mathcal{S}^{11}_m} [\min_{\theta^i,\gamma^i}\mC(y_{(\tau_{i-1}+1):\tau_i}|\theta^i,\gamma^i)+\iP^1]\right\}.
\end{align*}
Rearrange the order of optimization between $\theta^o$ and $\bftau$ in \eqref{MDL1} and \eqref{MDL2}, we have
\begin{align*}
F^*_o(n)&=\min_{\theta^o}\min_{\bftau \in \mT(n)}\left\{\sum_{i \in\mathcal{S}^{oo}_m} [\min_{\gamma^i}\mC(y_{(\tau_{i-1}+1):\tau_i}|\theta^o,\gamma^i)+\iP^o] + \sum_{i \in\mathcal{S}^{o1}_m} [\min_{\theta^i,\gamma^i}\mC(y_{(\tau_{i-1}+1):\tau_i}|\theta^i,\gamma^i)+\iP^1]\right\}\\
&=\min_{\theta^o} F^*_o(n;\theta^o),\\
F^*_1(n)&=\min_{\theta^o}\min_{\bftau \in \mT(n)}\left\{\sum_{i \in\mathcal{S}^{1o}_m} [\min_{\gamma^i}\mC(y_{(\tau_{i-1}+1):\tau_i}|\theta^o,\gamma^i)+\iP^o] + \sum_{i \in\mathcal{S}^{11}_m} [\min_{\theta^i,\gamma^i}\mC(y_{(\tau_{i-1}+1):\tau_i}|\theta^i,\gamma^i)+\iP^1]\right\}\\
&=\min_{\theta^o} F^*_1(n;\theta^o).
\end{align*}
Denote $F^*(n;\theta^o)=\min(F^*_o(n;\theta^o),F^*_1(n;\theta^o)),$ we have
\begin{align*}
F^*(n)=\min(F^*_o(n),F^*_1(n))=\min_{\theta^o} F^*(n;\theta^o).
\end{align*}
Thus if we can solve $F^*(n;\theta^o)$ efficiently for any given $\theta^o$ and $F^*(n;\theta^o)$ is further a smooth function of $\theta^o$, we can efficiently solve $F^*(n)$ in a profile-likelihood fashion. In the following two subsections, we describe the two-stage optimization procedure in detail.

\subsection{Alternating PELT under known normal state parameter $\theta^o$}
In this subsection, for a given $\theta^o$, we propose an efficient alternating dynamic programming algorithm for solving $F^*(n;\theta^o)=\min(F^*_o(n;\theta^o),F^*_1(n;\theta^o))$.

Denote $\mC^o(y_{(t+1):s})=\min_{\gamma}\mC(y_{(t+1):s}|\theta^o,\gamma)$ and $\mC^1(y_{(t+1):s})=\min_{\theta,\gamma}\mC(y_{(t+1):s}|\theta,\gamma)$. Under the epidemic change-point setting, a normal state is always followed by an epidemic state and vice versa. Thus, there is an implicit alternating recursion between $F^*_o(s;\theta^o)$ and $F^*_1(s;\theta^o)$ where
\begin{align}
\hspace{-0.35cm} F^*_o(s;\theta^o)=\min_{t<s}\{F^*_1(t;\theta^o)+\mC^o(y_{(t+1):s})+P^o \} \text{, } F^*_1(s;\theta^o)=\min_{t<s} \{F^*_o(t;\theta^o)+\mC^1(y_{(t+1):s})+P^1\}.
\label{alternating_relationship}
\end{align}

Equations \eqref{alternating_relationship} provides a recursive relationship between the minimum cost $F_o^*(s;\theta^o)$ for $y_{1:s}$ and the minimum cost $F_1^*(t;\theta^o)$ for $y_{1:t}$ with $t<s$, and similarly between $F_1^*(s;\theta^o)$ and $F_o^*(t;\theta^o)$. Thus, to obtain $F^*(n;\theta^o)=\min(F^*_o(n;\theta^o),F^*_1(n;\theta^o))$, we can solve $F_o^*(s;\theta^o)$ and $F_1^*(s;\theta^o)$ alternatingly by recursion for $s=1,2,\ldots,n.$ {We call this an alternating dynamic programming algorithm and its computational cost can be easily shown to be $O(n^2)$.}

To further reduce the computational cost, we propose an alternating PELT~(aPELT) algorithm by extending the pruning idea of PELT in \cite{Killick2012}. The key idea is that when calculating $F^*_o(s;\theta^o)$ and $F^*_1(s;\theta^o)$ via recursion \eqref{alternating_relationship}, we do not need to consider all $t<s$. Instead, we only need to consider a subset of $t<s$ by adding a pruning step. Theorem \ref{aPELTpruning} gives the pruning step for aPELT and provides its theoretical guarantee.

\begin{theorem}
\label{aPELTpruning}
Given $\theta^o,$ assume that there exists a constant $K_o$ such that for all $t<s<n$,
$$\mC^o(y_{(t+1):s}) + \mC^o(y_{(s+1):n}) + K_o \leq \mC^o(y_{(t+1):n}).$$
Then if
$$F^*_1(t;\theta^o) + \mC^o(y_{(t+1):s}) + K_o > F^*_1(s;\theta^o)$$
holds, at a future time $n>s$, $t$ can never be the optimal last change-point for $F^*_o(n;\theta^o)$ prior to $n$. Similarly, assume that there exists a constant $K_1$ such that for all $t<s<n$,
$$\mC^1(y_{(t+1):s}) + \mC^1(y_{(s+1):n}) + K_1 \leq \mC^1(y_{(t+1):n}).$$
Then if
$$F^*_o(t;\theta^o) + \mC^1(y_{(t+1):s}) + K_1 > F^*_o(s;\theta^o),$$
holds, at a future time $n>s$, $t$ can never be the optimal last change-point for $F^*_1(n;\theta^o)$ prior to $n$.
\end{theorem}

{Compared with PELT, the pruning scheme for aPELT is more complicated as aPELT needs to solve two optimization of $F_0^*(n;\theta^o)$ and $F_1^*(n;\theta^o)$ simultaneously, and thus requires two sets of pruning for $F_0^*(n;\theta^o)$ and $F_1^*(n;\theta^o)$ respectively.} An interesting phenomenon in Theorem \ref{aPELTpruning} is that the pruning for $F_0^*(n;\theta^o)$ requires the values of $F_1^*(s;\theta^o)$ and vice versa, which again calls for alternating optimization of $F_0^*(n;\theta^o)$ and $F_1^*(n;\theta^o)$. Based on Theorem \ref{aPELTpruning}, the pseudo-code of aPELT with known normal state parameter $\theta^o$ is given in Algorithm \ref{aPELT_algo} and we name it aPELT($\theta^o$).

\textbf{Remark 1} (pruning condition): If $\mC(y_{(t+1):s};\theta,\gamma)$ is the log-likelihood function of $y_{(t+1):s}$, it can be easily seen that the constant $K_o$ and $K_1$ in Theorem \ref{aPELTpruning} exist and can be set as 0 for any $\theta^o$.

{\textbf{Remark 2} (computational cost of aPELT): For a fixed $\theta^o$, aPELT$(\theta^o)$ essentially runs two pruned dynamic programming simultaneously in an alternating fashion. Thus, for any given $\theta^o$, the computational upper bound for solving $F^*(n;\theta^o)$ via aPELT$(\theta^o)$ is $O(n^2)$. Due to the pruning scheme, the computational performance of aPELT$(\theta^o)$ is expected to resemble PELT, which is more efficient with a growing number of change-points. As is demonstrated by extensive numerical experiments in Section 4.2, the computational cost of aPELT$(\theta^o)$ is comparable to PELT across a wide range of observation length $n$ regardless of the number of change-points $m$.}

\begin{algorithm}
	\caption{aPELT($\theta^o$): aPELT algorithm with known normal state parameter $\theta^o$}\label{aPELT_algo}
	\begin{algorithmic}[1]
	\BState \textbf{Initialize}: Set $F^*_o(0;\theta^o)=F^*_1(0;\theta^o)=0, cp^1(0)=cp^o(0)=\mbox{NULL}, R_1^o=R_1^1=\{0\}$.
	\BState \textbf{Iterate}: for $s=1,\ldots,n$
	\State\hspace{0.2cm} {[Alternating dynamic programming: (i)-(iii)]}
		\State \indent (i). Calculate  $F^*_o(s;\theta^o)=\min_{t \in R_{s}^o}\{F^*_1(t;\theta^o)+\mC^o(y_{(t+1):s})+P^o \}$,\\ \hspace{3.15cm}$F^*_1(s;\theta^o)=\min_{t \in R_{s}^1} \{F^*_o(t;\theta^o)+\mC^1(y_{(t+1):s})+P^1\}$.
		\State \indent (ii). Let $t^*_o=\argmin_{t \in R_{s}^o}\{F^*_1(t;\theta^o)+\mC^o(y_{(t+1):s})+P^o \}$,\\
		\hspace{2.25cm}$t^*_1=\argmin_{t \in R_{s}^1} \{F^*_o(t;\theta^o)+\mC^1(y_{(t+1):s})+P^1\}$.
		\State \indent (iii). Set $cp^o(s)=\{cp^1(t^*_o), t^*_o\}$ and $cp^1(s)=\{cp^o(t^*_1), t^*_1\}$.
	\State \hspace{0.2cm} {[Pruning: (iv)]}
		\State \indent (iv). Set $R_{s+1}^o=\{t\in R_s^o \cup \{s\} : F^*_1(t;\theta^o)+\mC^0(y_{(t+1):s})+K_o\leq F^*_1(s;\theta^o) \}$,\\
		 	\hspace{2.2cm} $R_{s+1}^1=\{t\in R_s^1 \cup \{s\} : F^*_o(t;\theta^o)+\mC^1(y_{(t+1):s})+K_1\leq F^*_o(s;\theta^o) \}.$
    \BState \textbf{Output}: If $F^*_o(n;\theta^o)\leq F^*_1(n;\theta^o)$, return $cp^o(n)$ and alternating states with last state normal.\\
                  \hspace{1.6cm}  Otherwise, return $cp^1(n)$ and alternating states with last state epidemic.
\end{algorithmic}
\end{algorithm}

\subsection{Alternating PELT under unknown normal state parameter $\theta^o$}
For many applications, the normal state parameter $\theta^o$ is naturally known and thus aPELT($\theta^o$) proposed in Section 3.2 is sufficient. For example, in DNA copy number variation, the mean log-ratio between the test and reference sequence is typically 0 when there is no variation; in multiple testing with locally clustered signals, the normal state is uniform distribution $U(0,1)$. Nevertheless, for the sake of generality, it is of interest to cover the case of unknown $\theta^o$. In this section, we discuss two extensions of aPELT, namely aPELT\_profile and aPELT\_plugin, to handle such situation.

\subsubsection{Profile aPELT}
The proposed aPELT($\theta^o$) in Algorithm \ref{aPELT_algo} can find the exact minimum of $F^*(n;\theta^o)$ for a given normal state parameter $\theta^o$, thus if $F^*(n;\theta^o)$ is a smooth function of $\theta^o$, we can solve $F^*(n)=\min_{\theta^o}F^*(n;\theta^o)$ by a standard optimization algorithm such as gradient descent. In addition, as a byproduct, $\theta^o$ can be estimated by $\tilde{\theta}^o=\argmin_{\theta^o}F^*(n;\theta^o)$. This two-stage procedure shares the same spirit as profile likelihood, thus we name it aPELT\_profile.

To justify aPELT\_profile, we investigate the behavior of $F^*(n;\theta^o)$ and show in Theorem \ref{piecewise_differentiable} that in general $F^*(n;\theta^o)$ is a piecewise smooth function of $\theta^o$, and thus can be solved by a gradient-based algorithm. By writing out $F^*_o(n;\theta^o)$ and $F^*_1(n;\theta^o)$ explicitly, we have
\begin{align*}
	F^*_o(n;\theta^o)=&\min_{\bftau \in \mT(n)}\left[\sum_{i \in\mathcal{S}^{oo}_m} [\min_{\gamma^i}\mC(y_{(\tau_{i-1}+1):\tau_i}|\theta^o,\gamma^i)+\iP^o] + \sum_{i \in\mathcal{S}^{o1}_m} [\min_{\theta^i,\gamma^i}\mC(y_{(\tau_{i-1}+1):\tau_i}|\theta^i,\gamma^i)+\iP^1]\right],\\
	F^*_1(n;\theta^o)=&\min_{\bftau \in \mT(n)}\left[\sum_{i \in\mathcal{S}^{1o}_m} [\min_{\gamma^i}\mC(y_{(\tau_{i-1}+1):\tau_i}|\theta^o,\gamma^i)+\iP^o] + \sum_{i \in\mathcal{S}^{11}_m} [\min_{\theta^i,\gamma^i}\mC(y_{(\tau_{i-1}+1):\tau_i}|\theta^i,\gamma^i)+\iP^1]\right].
\end{align*}

For a given change-point configuration $\bftau$, we denote
$C_1(\bftau)=\sum_{i \in\mathcal{S}^{o1}_m} [\min_{\theta^i,\gamma^i}\mC(y_{(\tau_{i-1}+1):\tau_i}|\theta^i,\gamma^i)+\iP^1]$ and $C_2(\bftau)=\sum_{i \in\mathcal{S}^{11}_m} [\min_{\theta^i,\gamma^i}\mC(y_{(\tau_{i-1}+1):\tau_i}|\theta^i,\gamma^i)+\iP^1]$, as the two quantities are constants not depending on $\bftau$. We further denote
$g_1(\theta^o;\bftau)=\sum_{i \in\mathcal{S}^{oo}_m} [\min_{\gamma^i}\mC(y_{(\tau_{i-1}+1):\tau_i}|\theta^o,\gamma^i)+\iP^o]$ and 
$g_2(\theta^o;\bftau)=\sum_{i \in\mathcal{S}^{1o}_m} [\min_{\gamma^i}\mC(y_{(\tau_{i-1}+1):\tau_i}|\theta^o,\gamma^i)+\iP^o]$, as the two quantities are functions of $\theta^o$. Therefore, we have
\begin{align*}
F^*(n;\theta^o)=\min(F^*_o(n;\theta^o),F^*_1(n;\theta^o))=\min&\left\{\min_{\bftau \in \mT(n)}\left[g_1(\theta^o;\bftau)+C_1(\bftau)\right],\min_{\bftau \in \mT(n)}\left[g_2(\theta^o;\bftau)+C_2(\bftau)\right]\right\}.
\end{align*}
In other words, $F^*(n;\theta^o)$ is the minimum of 2$|\mT(n)|$ functions of $\theta^o$, where $|\cdot|$ denotes the cardinality of a set. Intuitively, if for each $\bftau$, $g_1(\theta^o;\bftau)$ and $g_2(\theta^o;\bftau)$ are smooth functions of $\theta^o$, $F^*(n;\theta^o)$ should also be a (piecewise) smooth function of $\theta^o$. This notion is later formalized in Theorem \ref{piecewise_differentiable}. 

Denote $\Theta\in \mathbb{R}^{p_1}$ as the parameter space of the normal state parameter $\theta^o$ and denote $\mathring{\Theta}$ as the interior of $\Theta$. Before stating Theorem \ref{piecewise_differentiable}, we first state two assumptions on the behavior of the $2|\mT(n)|$ functions in $\big\{g_1(\theta^o;\bftau),g_2(\theta^o;\bftau)|\bftau \in \mT(n)\big\}$.

\begin{assumption}[Smoothness]
	\label{continuity}
	Every function in $\big\{g_1(\theta^o;\bftau),g_2(\theta^o;\bftau)|\bftau \in \mT(n)\big\}$ is a differentiable function of $\theta^o$ and has a unique global minimizer in $\mathring{\Theta}$. WLOG, further assume that the global minimizers and the minimum values of different functions are different. 
\end{assumption}

\begin{assumption}[Finite Partition]
	\label{partition}
	There exists a finite partition $\{\Theta_1,\ldots,\Theta_{N(n)}\}$ of $\Theta$ where each $\Theta_i$ is a connected set in $\mathbb{R}^p$ and there is no intersection among functions in $\big\{g_1(\theta^o;\bftau),g_2(\theta^o;\bftau)|\bftau \in \mT(n)\big\}$ in each interior set $\mathring{\Theta}_i$, for $i=1,\ldots,N(n)$. 
\end{assumption}

Both Assumptions \ref{continuity} and \ref{partition} are mild and are expected to hold for common loss functions $\mC(y_{(t+1):s};\theta,\gamma)$ such as log-likelihood. In Section 3.5, we verify Assumptions \ref{continuity}-\ref{partition} for some classical change-point settings. Assumption \ref{partition} is used to evoke intermediate value theorem in the proof and show that $F^*(n;\theta^o)$ is \textit{piecewise} differentiable on $\Theta.$ {As an intuitive example, for $\Theta=(\theta_L, \theta_U) \subseteq \mathbb{R}$, a sufficient condition for Assumption \ref{partition} is that the functions in $\big\{g_1(\theta^o;\bftau),g_2(\theta^o;\bftau)|\bftau \in \mT(n)\big\}$ have finite intersection points at, say $\theta_1<\theta_2<\ldots <\theta_{N(n)-1}$. Thus, the finite partition in Assumption 2 can be set as $\bigcup_{i=1}^{N(n)}\Theta_i=\bigcup_{i=1}^{N(n)}[\theta_{i-1},\theta_{i}]$, where we define $\theta_0=\theta_L$ and $\theta_{N(n)}=\theta_U$.}

\begin{theorem}
	\label{piecewise_differentiable}
	Under Assumption \ref{continuity}, $F^*(n;\theta^o)$ is a continuous function of $\theta^o$ and has a unique global minimizer $\theta^{o*}$ in $\mathring{\Theta}$, and there exists an open neighborhood $\mathcal{N}(\theta^{o*})$ of $\theta^{o*}$ such that $F^*(n;\theta^o)$ is differentiable on $\mathcal{N}(\theta^{o*})$. If in addition Assumption \ref{partition} holds, then we further have that $F^*(n;\theta^o)$ is differentiable in every $\mathring{\Theta}_i$ and $\theta^{o*} \in \mathring{\Theta}_i$ for some $i=1,\ldots,N(n)$.
\end{theorem}

By Theorem \ref{piecewise_differentiable}, under Assumption \ref{continuity}, with a properly chosen starting point, combined with a standard optimization algorithm such as gradient descent, aPELT\_profile can successfully locate the global minimizer of $F^*(n;\theta^o)$, and thus simultaneously estimate the unknown normal state parameter $\theta^o$ and the unknown change-points $\bftau^o.$ With the additional Assumption \ref{partition}, we know that $F^*(n;\theta^o)$ is piecewise differentiable on $\Theta$ and is differentiable in every $\mathring{\Theta}_i$, which further justifies the use of gradient-based optimization such as gradient descent. Based on Theorem \ref{piecewise_differentiable}, the pseudo-code of aPELT\_profile with unknown normal state parameter is given in Algorithm \ref{aPELT_algo2}.

\begin{algorithm}
	\caption{aPELT\_profile: aPELT algorithm with unknown normal state parameter\label{aPELT_algo2}}
		\begin{algorithmic}[1]
			\BState Given an initial point $\theta^o_0$
			\BState Run gradient descent on $F^*(n;\theta^o)$ starting from $\theta^o_0$, where the function value of $F^*(n;\theta^o)$ is evaluated via aPELT($\theta^o$).
			\BState \textbf{Output}: Return $\tilde{\theta}^o=\argmin_{\theta^o}F^*(n;\theta^o)$ given by gradient descent and output of aPELT($\tilde{\theta}^o$).
		\end{algorithmic}
\end{algorithm}

In practice, for the choice of the initial point $\theta^o$, we can either select a set of different starting points across $\Theta$ and run aPELT\_profile in parallel or we can initiate aPELT\_profile from a reasonable estimator $\hat{\theta}^o$ of $\theta^o$ as is discussed in the following Section 3.3.2.

{The computational performance of aPELT\_profile further depends on the smoothness of $F^*(n;\theta^o)$ and the second-stage optimization. It is trivial to see that the computational cost of aPELT\_profile is $d$ times that of aPELT$(\theta^o)$, where $d$ is the number of function evaluations performed during the second-stage optimization. We empirically investigate the computational performance of aPELT\_profile in Section 4.2 through extensive numerical experiments.}

\subsubsection{Plug-in aPELT}
When the normal state parameter $\theta^o$ is unknown, another natural way to proceed is to first obtain an estimated $\hat{\theta}^o$ and then run aPELT($\hat{\theta}^o$) as if it is the true parameter. We call this method aPELT\_plugin. As expected, the performance of aPELT\_plugin is closely related to the accuracy of $\hat{\theta}^o$. With an accurate estimator for ${\theta}^o$, aPELT\_plugin should have decent performance.

The estimator $\hat{\theta}^o$ needs be chosen based on specific cases. For example, if $y_{1:n}$ is a sequence of univariate Gaussian random variables with epidemic mean change and a normal state mean $\mu^o$, then one possible estimator $\hat{\mu}^o$ is the median of estimated mean from a sequence of short screening-windows. This should work well if the epidemic state lasts shorter than the normal state.

\subsection{ Examples of aPELT for epidemic mean change and epidemic variance change}
In this section, we provide detailed examples of aPELT for epidemic change-point detection in mean or variance of a sequence of independent Gaussian random variables. Consider a sequence of observations $y_{1:n}$ such that 
\begin{align}
\label{mean change}
y_t=\mu_t+\varepsilon_t,
\end{align}
where $\{\varepsilon_t\}$ is a sequence of independent Gaussian noise following $N(0,\sigma_t^2)$. We consider two cases: (a).\ There are epidemic changes in mean $\mu_{1:n}$ with unknown variance, and (b).\ There are epidemic changes in variance $\sigma^2_{1:n}$ with unknown mean. For epidemic mean change, $\mu$ is the parameter of interest and $\sigma^2$ is the nuisance parameter, while for epidemic variance change, $\sigma^2$ is the parameter of interest and $\mu$ is the nuisance parameter.

{Note that due to its model selection nature, aPELT allows the nuisance parameter to be homogeneous or to simultaneously change with the parameter of interest at the same change-points, and thus has a broader applicability than testing-based change-point detection methods. We discuss more about this point in Section \ref{subsec:new_aPELT}.}

For both cases, we set $\mC(y_{(t+1):s}|\mu,\sigma)$ to be twice the negative log-likelihood of Gaussian random variables, where
\begin{align}
\label{GaussianLikelihood}
\mC(y_{(t+1):s}|\mu,\sigma)={(s-t)}\log(2\pi\sigma^2)+{\sum_{i=t+1}^{s}(y_i-\mu)^2}/{\sigma^2}.
\end{align}

We employ BIC for model penalty and set $P^o=2\log n$ and $P^1=3\log n$ since for an extra epidemic state, we need to record the location of the change-point and two parameters $(\mu,\sigma)$, while for an extra normal state, we do not need to record the common normal state parameter.

Proposition \ref{piecewise_continuous} characterizes the behavior of $F^*(n;\mu^o)$ and $F^*(n;\sigma^o)$ as functions of $\mu^o$ and $\sigma^o.$

\begin{proposition}
	\label{piecewise_continuous}
	For epidemic change-point detection in mean $\mu$ via \eqref{GaussianLikelihood}, Assumptions \ref{continuity} and \ref{partition} hold for $F^*(n;\mu^o)$. Thus, $F^*(n;\mu^o)$ is a continuous and piecewise differentiable function of $\mu^o$ and $F^*(n;\mu^o)$ has a global minimizer $\mu^{o*}$. More specifically, there exists a finite number $N=N(n)$ and a sequence $-\infty = \mu_0 <\mu_1<\cdots<\mu_{N+1}=\infty$ such that $F^*(n;\mu^o)$ is differentiable on each $(\mu_i,\mu_{i+1})$ and $\mu^{o*}$ is an interior point of one of the intervals. Moreover, the same result holds for $F^*(n;\sigma^o)$ of epidemic change-point detection in variance.
\end{proposition}

{In Figure S.2 of the supplement, we plot several typical realizations of $F^*(n;\mu^o)$ and discuss its optimization result, where the plots further confirm the continuous and piecewise differentiable claim in Proposition \ref{piecewise_continuous}.} Moreover, following the same argument, it is easy to prove that similar results as the one in Proposition \ref{piecewise_continuous} hold for other distributions from the exponential family, such as Poisson, Binomial and Exponential distributions, {further broadening the application of aPELT.}

{ \subsubsection{Modified aPELT for epidemic mean change with homoscedasticity}\label{subsec:new_aPELT}
A common assumption in the literature for mean change under \eqref{mean change} is homoscedasticity, where it is assumed that $\{\varepsilon_t\}$ is a sequence of \textit{i.i.d.}\ Gaussian noise following $N(0,\sigma^2)$. In other words, the change only happens in mean $\{\mu_t\}_{t=1}^n$ and the error term $\{\varepsilon_t\}$ is homoscedastic.

The homoscedasticity assumption is typically required by testing-based mean change detection algorithms~\citep[see][]{Olshen2004,Niu2012,FRYZLEWICZ2014,Shin2020} and is found to be reasonable for applications in DNA copy number variation. In practice, the variance $\sigma^2$ is unknown and needs to be replaced by an estimate $\hat{\sigma}_n^2$, where $\hat{\sigma}_n^2$ can be the median absolute deviation estimator~\citep{FRYZLEWICZ2014} or the local regression estimator~\citep{Niu2012}. 

Note that homoscedasticity is not intrinsically built into aPELT as \eqref{GaussianLikelihood} allows different segments to have different variance. To incorporate the homoscedasticity into aPELT, a simple strategy is to substitute $\sigma^2$ in \eqref{GaussianLikelihood} with an estimate $\hat{\sigma}_n^2$ such that $$\mC(y_{(t+1):s}|\mu,\hat{\sigma}_n)={(s-t)}\log(2\pi\hat{\sigma}_n^2)+{\sum_{i=t+1}^{s}(y_i-\mu)^2}/{\hat{\sigma}_n^2}.$$
The BIC penalty can thus be adjusted to $P^o=\log n$ and $P^1=2\log n$ as variance is fixed at $\hat{\sigma}_n^2$ across all segments. Intuitively, this modification can help increase the power of aPELT (when the homoscedasticity assumption is true) as the BIC penalty is lowered due to homoscedasticity. We refer to this variant as aPELT$_H$ and examine its performance for detecting DNA copy number variation via simulation in Section 4.2 and real data analysis in Section 5.2, where favorable performance over existing methods is observed.}

\section{Simulation Study}
In this section, we conduct extensive numerical experiments to compare the performance of aPELT with state-of-the-art change-point detection algorithms in the literature under various simulation settings for a univariate sequence $y_{1:n}$ generated via the framework \eqref{mean change}.

{Section 4.1 focuses on comparison between aPELT and PELT, as both methods are model selection-based change-point detection algorithms that can handle both mean change and variance change. Section 4.2 focuses on epidemic mean change with homoscedasticity and short epidemic states, which resembles the characteristics of real data from DNA copy number variation, and we compare aPELT with testing-based change-point detection algorithms designed for mean change, which are CBS in \cite{Olshen2004}, WBS in \cite{FRYZLEWICZ2014}, modSaRa in \cite{Xiao2015} and BWD in \cite{Shin2020}.}

\subsection{ Epidemic mean and variance change (aPELT v.s.\ PELT)}
In this section, we compare the performance of the three aPELT variants~(aPELT$(\theta^o)$, aPELT\_plugin and aPELT\_profile) with PELT in \cite{Killick2012} for epidemic mean change and epidemic variance change, {and demonstrate the advantage of incorporating the epidemic structure}. We present the detailed numerical result for epidemic mean change. The result for epidemic variance change is similar and can be found in Section $\S$4 of the supplementary material.

\textbf{Implementation details}: For PELT, we set $\mC(y_{(s+1):t};\mu,\sigma)$ as the Gaussian log-likelihood function \eqref{GaussianLikelihood} with unknown $(\mu,\sigma)$. For aPELT$(\theta^o)$, we assume the normal state parameter $\mu^o$ is known. For aPELT\_plugin, we estimate the normal state parameter $\mu^o$ by the median of sample mean from a sequence of screening-windows of size 10, i.e. $\hat{\mu}^o=\text{median}\{\text{mean}(y_{t+1:t+10}),t=0,\ldots,n-10\}$. Under the typical scenario where the sequence $y_{1:n}$ is mostly at the normal state, one would expect $\hat{\mu}^o$ to be an accurate estimator for $\mu^o.$ For aPELT\_profile, we set the starting point of optimization for $\mu^o$ at the estimated $\hat{\mu}^o$ as in aPELT\_plugin. For the aPELT methods, we employ BIC penalty with $P^o=2\log(n)$ for the normal state and $P^1=3\log(n)$ for the epidemic state. For PELT, the BIC penalty is $P=3\log(n)$.

\textbf{Data generating process}: Denote $m$ as the number of true change-points. WLOG, we set $m$ to be an odd number. For the location of the true change-points $\bftau^o$ and the alternating states, we first partition the sequence $y_{1:n}$ into $(m+1)/2$ equal-length segments, then for each of the $(m+1)/2$ segments, we further divide it into a normal state and an epidemic state, where the ratio between the length of the epidemic state and the normal state is randomly sampled from a uniform distribution $U(0.2,0.5)$, i.e. the normal state lasts longer than the epidemic state. Under this simulation setting, $y_{1:n}$ starts with the normal state and ends with an epidemic state. Based on \eqref{mean change}, we have $y_t=\mu_t+\varepsilon_t.$ For the mean structure $\{\mu_t\}$, we set $\mu^i$ of the $i$th segment as
\begin{align*}
\mu^i=\begin{cases} 
\mu^o=0 & \text{ if $i$th segment is at the normal state}, \\
(1-2 {B}_i(0.5))\cdot {U}_i(1,1.25) & \text{ if $i$th segment is at the epidemic state},
\end{cases}
\end{align*}
where ${B}_i(0.5)$ follows \textit{i.i.d.}\ Bernoulli distribution and $U_i(1,1.25)$ follows \textit{i.i.d.}\ uniform distribution. For the error structure $\{\varepsilon_t\}$, we consider two scenarios: (A) $\{\varepsilon_t\}$ are \textit{i.i.d.}\ errors with $\sigma_t\equiv1$, (B) $\{\varepsilon_t\}$ are independent errors with $\sigma_t=1.5$ for the normal state and $\sigma_t=1$ for the epidemic state. {The nuisance parameter~(variance) is homogeneous in Scenario (A) and heterogeneous in Scenario (B).} In Section \S3.2 of the supplementary material, we further examine the robustness of the algorithms against model misspecification under three more scenarios~(temporal dependence in error structure, long and short sinusoidal trends in mean structure), where aPELT consistently offers the best performance.

\textbf{Error measures}: To evaluate the performance of change-point estimation, following \cite{Killick2012}, we report the true positive rate~(TPR) and false positive rate~(FPR) of each algorithm. For each simulated sequence $y_{1:n}$, we define
\begin{align*}
&\text{TPR} = \frac{\text{number of correctly detected true change-point (CP)}}{m}\,,\\
&\text{FPR} = \frac{\text{total number of detected CP}-\text{number of correctly detected true CP}}{\text{total number of detected CP}}\,,
\end{align*}
where for a true change-point $\tau_i^o \in \bftau^o$, we consider it to be correctly detected if there is an estimated change-point $\hat{\tau}_{i'}$ within a distance of {$10$} points from $\tau_i^o$. To evaluate the performance of parameter estimation, we report the MSE of the estimated parameter as $\text{MSE}(\theta)=\left({\sum_{t=1}^{n}(\hat{\theta}_t-\theta_t)^2} \Big/ {n}\right)^{1/2}\,,$ where $\theta_{1:n}$ denotes the true parameter of interest and $\theta=\mu$ for epidemic mean change.

\textbf{Estimation performance}: We set $n=1000, 2000, 5000, 10000$ and consider $m=\sqrt{n/10}-1=9,13,21,31$ or $m=n/100-1=9,19,49,99$, i.e. the number of change-points $m$ grows sublinearly or linearly with $n$. For each simulation setting, we repeat the experiments 1000 times and report the average result. Note that the locations of true change-points $\bftau^o$ and alternating mean structures $\{\mu^{i}\}_{i=1}^{m+1}$ are simulated separately for each experiment. 

\begin{figure}[H]
	\begin{subfigure}{1\textwidth}
		\includegraphics[width=1.1\textwidth]{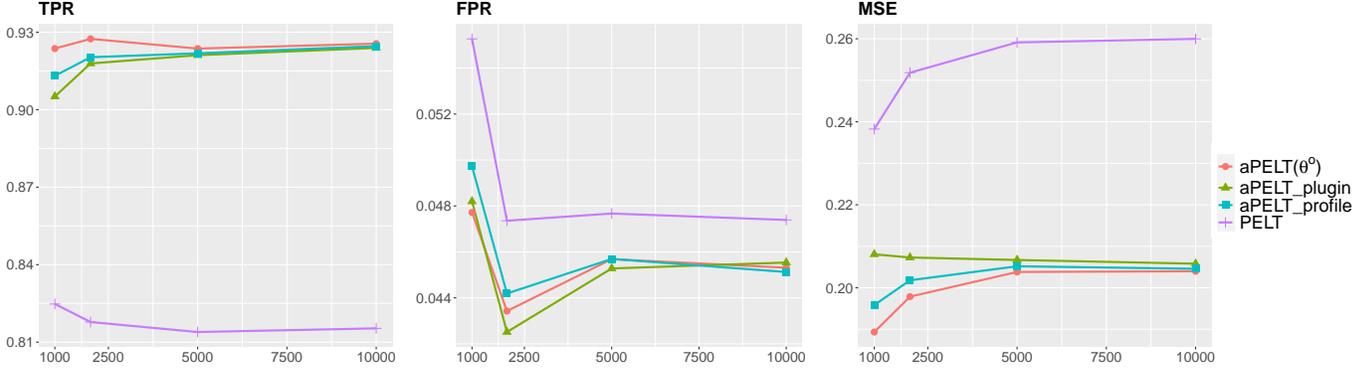}
		\caption{ Linearly growing $m$ + Scenario A~(IID error)}
	\end{subfigure}
	~
	\begin{subfigure}{1\textwidth}
		\includegraphics[width=1.1\textwidth]{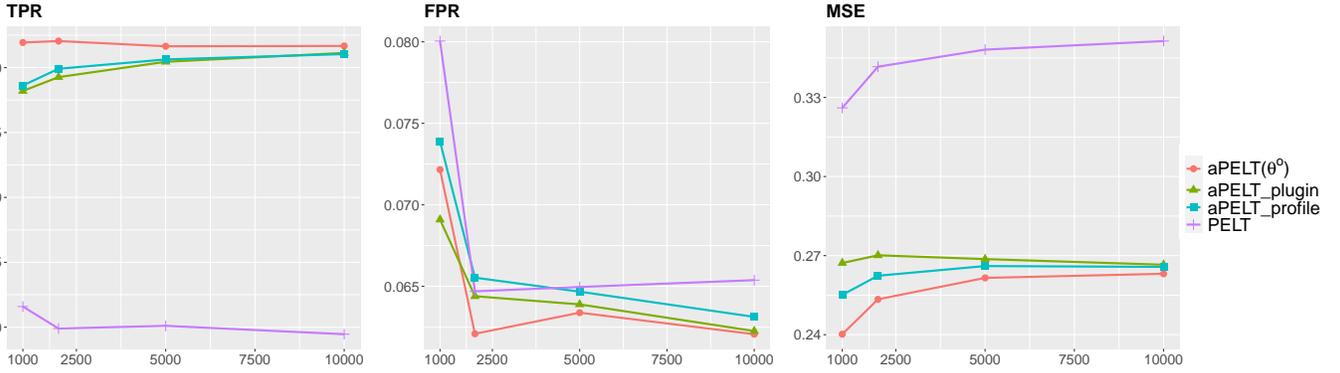}
		\caption{ Linearly growing $m$ + Scenario B~(Independent error with additional variance change)}
	\end{subfigure}
	~
	\begin{subfigure}{1\textwidth}
		\includegraphics[width=1.1\textwidth]{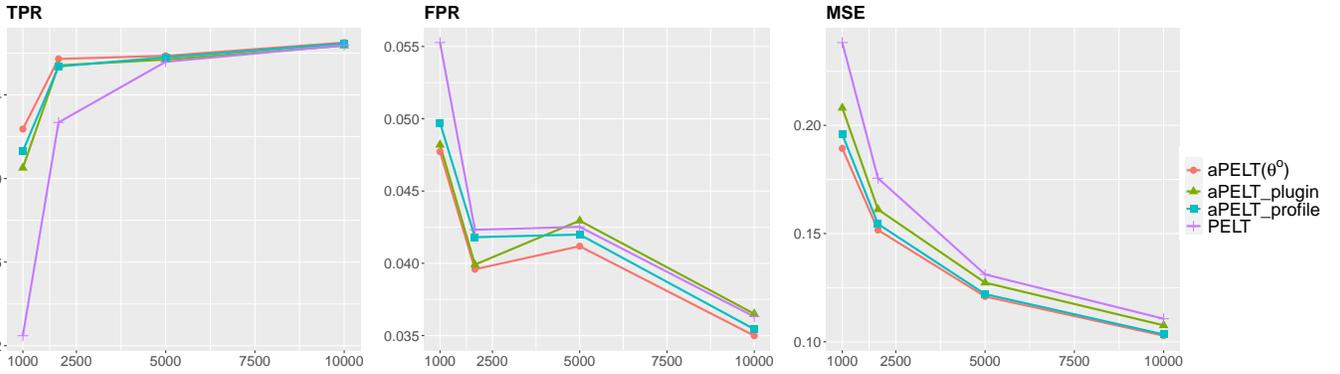}
		\caption{Sublinearly growing $m$ + Scenario A~(IID error)}
	\end{subfigure}
	~
	\begin{subfigure}{1\textwidth}
		\includegraphics[width=1.1\textwidth]{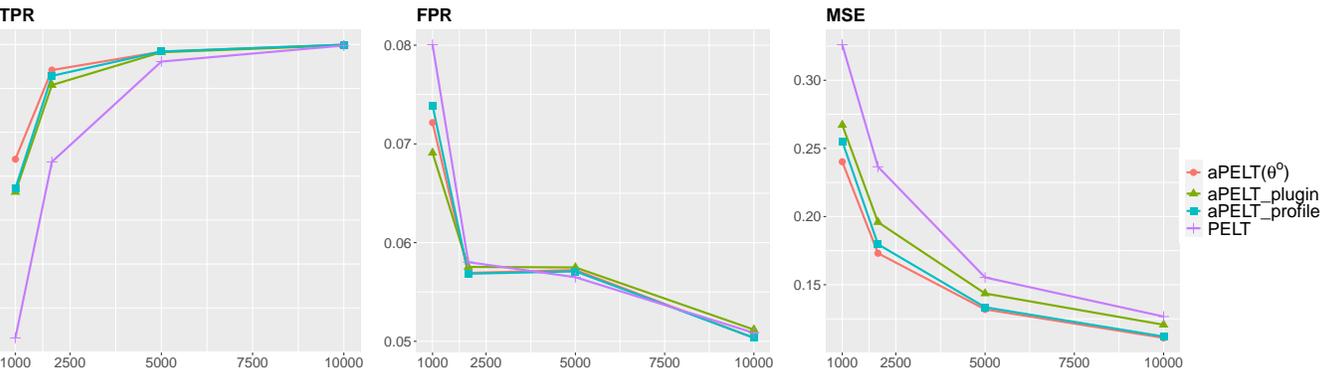}
		\caption{Sublinearly growing $m$ + Scenario B~(Independent error with additional variance change)}
	\end{subfigure}
	\caption{Performance of PELT and aPELT for epidemic mean change.}
	\label{epidemicMean}
\end{figure}

{Figure \ref{epidemicMean}(a)-(b) reports the estimation accuracy~(TPR, FPR and MSE) for linearly growing $m$. In general, the three aPELT algorithms offer the best performance, where aPELT$(\theta^o)$ performs the best, and aPELT\_plugin and aPELT\_profile give similar and slightly worse result. For both Scenarios (A) and (B), all variants of aPELT offer notably better performance than PELT in terms of TPR and MSE while maintaining smaller FPR. This further confirms the intuition that aPELT gains statistical efficiency by explicitly utilizing the alternating structure of the epidemic change.}

Figure \ref{epidemicMean}(c)-(d) compares the performance for sublinearly growing $m$. {Note that the performance of all algorithms improves with $n$.} Again, aPELT based algorithms offer notably better performance than PELT. One difference is that under the sublinear $m$, when sample size is large~($n>5000$), PELT catches up and gives comparable~(though slightly inferior) performance as aPELT.

\textbf{Optimization performance}: To conserve space, the optimization performance of aPELT\_profile w.r.t.\ function $F^*(n;\mu^o)$ is reported in Section \S3.1 of the supplementary material. In general, aPELT\_profile with a starting point at $\hat{\mu}^o$ can find the global minimizer with a reasonable success rate. For more intuition, Figure S.2 in the supplement further gives example plots of $F^*(n;\mu^o)$, where it is observed that $F^*(n;\mu^o)$ is indeed piecewise smooth and whether aPELT\_profile stops at the global minimizer largely depends on the starting point $\hat{\mu}^o.$ Thus, one can start the optimization at a set of initial values of $\mu^o$ to ensure the recovery of global minimizer.

\textbf{Computation performance}: To investigate the computational cost of aPELT, we run additional simulations for PELT and the three aPELT algorithms with number of change-points $m=n/100$, $m=\sqrt{n/10}$ and $m=10$, where $n=1000,2000,5000,10000,20000,30000,40000,50000.$ 

\begin{figure}[h]
	\includegraphics[width=1.1\textwidth]{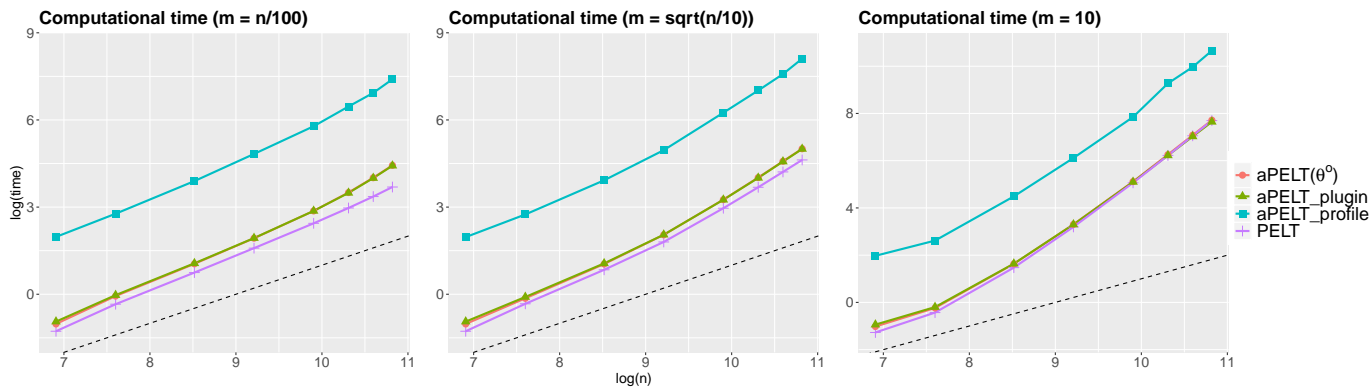}
	\caption{Log average computational time~(in seconds) \textit{v.s.} Log observation length $n$ for aPELT($\theta^o$), aPELT\_plugin, aPELT\_profile and PELT. The reference line (dashed line) is $y=-9+x$.}
	\label{epidemicMean_time_log}
\end{figure}

Figure \ref{epidemicMean_time_log} plots the average computational time across 1000 experiments \textit{v.s.} the observation length $n$ in log-log scale. As can be seen, aPELT has relatively higher~(but comparable) computational cost than PELT. In line with the theory in \cite{Killick2012}, the pruning procedure of aPELT and PELT is most effective when the number of change-points $m$ grows linearly with $n$. Empirically, the computational time of the four algorithms roughly increases at the same order with $n$ under all three schemes of change-point number $m$. As expected, aPELT\_profile incurs the highest computational cost~(around $10\sim 20$ times that of aPELT($\theta^o$)).

\subsection{{Epidemic mean change with homoscedasticity}}
{ In this section, we focus on epidemic mean change with homoscedasticity and short epidemic states, which resembles the characteristics of real data from DNA copy number variation~(CNV). We compare aPELT$_H$ proposed in Section \ref{subsec:new_aPELT} with CBS in \cite{Olshen2004}, WBS in \cite{FRYZLEWICZ2014}, modSaRa in \cite{Xiao2015} and BWD in \cite{Shin2020}.

The four competing methods are testing-based detection algorithms for mean change and require homoscedasticity. CBS and modSaRa are widely used methods for CNV detection in the literature. WBS is a modified binary segmentation with improved power for short segments. BWD is a novel CNV detection algorithm based on a backward principle tailored for detecting short signals. 

\textbf{Implementation details}: For CBS, WBS, modSaRa and BWD, we use the default setting in R packages \texttt{DNAcopy}, \texttt{wbs}, \texttt{modSaRa} and \texttt{bwd} respectively. For BWD, we implement the modified BWD in \cite{Shin2020}, which utilizes the information of normal state mean $\mu^o$ and achieves better performance. For aPELT$_H$, the unknown variance $\hat{\sigma}_n^2$ is estimated via local regression as in modSaRa and BWD, where $\hat{\sigma}_n^2=n^{-1}\sum_{i=1}^{n}(y_i-\bar{y}_i^{(h)})^2$ with $\bar{y}_i^{(h)}=\sum_{j=i-h}^{i+h}y_j/(2h+1)$ for a window size $h=10$, and we assume the normal state mean $\mu^o$ is known. As argued in \cite{Shin2020}, this is a reasonable assumption as for CNV detection, $\mu^o$ is typically known as 0.

\textbf{Data generating process}: The simulation setting is largely borrowed from \cite{Shin2020}. Denote $\mathbf{1}_{\{\cdot\}}$ as the indicator function, $y_{1:n}$ is simulated via a special case of \eqref{mean change} such that
$$y_t=\sum_{k=1}^{K}\delta \mathbf{1}_{\{t\in I_k\}}+\varepsilon_t,$$
where $\{\varepsilon_t\}$ is \textit{i.i.d.}\ $N(0,\sigma^2)$, and $\{I_k\}_{k=1}^K$ are $K$ intervals indicating the epidemic states with mean $\delta$ and the normal states are $\{1\text{:}n\}\setminus \{I_k\}_{k=1}^K$ with mean $\mu^o=0$. We set the number of epidemic segments~(i.e.\ true CNV signals) $K=n/1000+1$ and $I_k=\{(\lfloor kn/K \rfloor-L+1): \lfloor kn/K \rfloor \}$ for $k=1,\cdots,K$. Thus each epidemic segment is of length $L.$ As a concrete example, for $n=1000, L=5$, we have $K=2$, $I_1=\{496\text{:}500\}$ and $I_2=\{996\text{:}1000\}$.

\textbf{Error measures}: The estimated change-points $(\hat{\tau}_1,\cdots,\hat{\tau}_{\hat{m}})$ partition $y_{1:n}$ into $\hat{m}+1$ segments $\hat{I}_k=\{(\hat{\tau}_k+1):\hat{\tau}_{k+1}\}, k=0,1,\cdots,\hat{m}$ where by convention $\hat{\tau}_0=0$ and $\hat{\tau}_{\hat{m}+1}=n$. Following \cite{Shin2020}, we count $\hat{I}_k$ as a detected CNV signal if $|\hat{I}_k|<2L$ and we consider an epidemic segment $I_k$~(true signal) as correctly detected if there exists a detected signal $\hat{I}_{k'}$ such that $\hat{I}_{k'}\cap I_k \neq \varnothing$. We use the two measures defined in \cite{Shin2020} to assess the performance of methods:
\begin{align*}
&\text{Sensitivity} = \text{(number of correctly detected signals)}~/~(\text{number of true signals}, K)\,,\\
&\text{Precision} = \text{(number of correctly detected signals)}~/~(\text{number of detected signals})\,.
\end{align*}
Sensitivity measures the ability to detect true signals and precision measures reliability of the detected signals. Note that both measures range between zero and one and a method is perfect if both measures have a value of one. 

\textbf{Estimation performance}: Same as in \cite{Shin2020}, we set $(n,L,\delta)=\{1000,3000,5000\}\times \{5,10\}\times \{1.5,2,2.5\}$ with $\sigma=1$ for the homoscedastic variance. Note that the length $L$ of the epidemic segment is rather short, which resembles the case in CNV detection and makes its discovery challenging. For each simulation setting, we repeat the experiments 1000 times.

Table \ref{short_segment} reports the average sensitivity and precision for each method. In general, aPELT$_H$ and BWD provide the best performance. For $n=1000$, aPELT$_H$ indeed gives the best sensitivity and (almost the best) precision. For $n=3000,5000$, the epidemic signal gets further diluted and BWD offers the best sensitivity due to its novel backward principle, while aPELT$_H$ gives the second highest sensitivity with a more reliable precision. Remarkably, compared to CBS, which is the most widely used CNV detection algorithm in the literature, aPELT$_H$ gives better performance in both sensitivity and precision under all simulation settings. In summary, the numerical result clearly demonstrates the efficiency and promising potential of aPELT$_H$ for CNV detection, which is further confirmed by the real data analysis in Section 5.2.

\begin{table}[ht]
	\centering
	\centerline{\small 
	\begin{tabular}{ccrrrrrrr|rrrrrrr}
		\hline\hline
		& $L$ & \multicolumn{7}{c|}{5} & \multicolumn{7}{c}{10}  \\
		& $\delta$ & \multicolumn{2}{c}{1.5} & \multicolumn{2}{c}{2.0} & \multicolumn{2}{c}{2.5} & & \multicolumn{2}{c}{1.5} & \multicolumn{2}{c}{2.0} & \multicolumn{2}{c}{2.5} & \\
		$n$ & Methods & Sen. & Pre. & Sen. & Pre. & Sen. & Pre. & Time & Sen. & Pre. & Sen. & Pre. & Sen. & Pre. & Time \\ 
		\hline
		& CBS & 0.087 & 0.956 & 0.392 & 0.990 & 0.787 & 0.993 & 0.10 & 0.543 & 0.974 & 0.932 & 0.987 & 1.000 & 0.991 & 0.05 \\ 
		& WBS & 0.114 & 0.975 & 0.438 & 0.995 & 0.801 & 0.997 & 0.06 & 0.559 & 0.996 & 0.932 & 0.996 & 1.000 & 0.998 & 0.06 \\ 
		1000 & modSaRa & 0.130 & 0.805 & 0.311 & 0.947 & 0.449 & 0.983 & 1.21 & 0.493 & 0.962 & 0.771 & 0.980 & 0.888 & 0.981 & 1.23 \\ 
		& BWD & 0.241 & 0.910 & 0.603 & 0.965 & 0.867 & 0.981 & 0.03 & 0.711 & 0.962 & 0.967 & 0.977 & 0.998 & 0.984 & 0.03 \\ 
		& aPELT$_H$ & 0.283 & 0.970 & 0.660 & 0.989 & 0.913 & 0.992 & 0.63 & 0.750 & 0.983 & 0.978 & 0.989 & 1.000 & 0.990 & 0.49 \\ \hline
		& CBS & 0.056 & 0.966 & 0.316 & 0.986 & 0.747 & 0.991 & 0.35 & 0.479 & 0.983 & 0.935 & 0.987 & 0.998 & 0.988 & 0.16 \\ 
		& WBS & 0.050 & 0.968 & 0.271 & 0.996 & 0.669 & 1.000 & 0.11 & 0.396 & 0.998 & 0.872 & 1.000 & 0.996 & 1.000 & 0.11 \\ 
		3000 & modSaRa & 0.133 & 0.865 & 0.309 & 0.968 & 0.507 & 0.984 & 3.58 & 0.430 & 0.969 & 0.818 & 0.978 & 0.944 & 0.977 & 3.54 \\ 
		& BWD & 0.165 & 0.900 & 0.544 & 0.967 & 0.861 & 0.984 & 0.11 & 0.656 & 0.968 & 0.958 & 0.979 & 0.997 & 0.989 & 0.11 \\ 
		& aPELT$_H$ & 0.138 & 0.977 & 0.460 & 0.994 & 0.827 & 0.997 & 6.24 & 0.596 & 0.993 & 0.951 & 0.996 & 0.998 & 0.997 & 3.23 \\ \hline
		& CBS & 0.050 & 0.979 & 0.301 & 0.994 & 0.729 & 0.994 & 0.64 & 0.454 & 0.990 & 0.923 & 0.991 & 0.997 & 0.992 & 0.27 \\ 
		& WBS & 0.032 & 0.994 & 0.206 & 0.999 & 0.553 & 1.000 & 0.16 & 0.310 & 0.998 & 0.813 & 1.000 & 0.983 & 1.000 & 0.17 \\ 
		5000 & modSaRa & 0.126 & 0.892 & 0.261 & 0.971 & 0.477 & 0.990 & 5.94 & 0.352 & 0.968 & 0.841 & 0.981 & 0.964 & 0.978 & 5.89 \\ 
		& BWD & 0.154 & 0.921 & 0.504 & 0.974 & 0.844 & 0.986 & 0.20 & 0.622 & 0.972 & 0.947 & 0.983 & 0.997 & 0.989 & 0.19 \\ 
		& aPELT$_H$ & 0.096 & 0.987 & 0.385 & 0.998 & 0.764 & 0.999 & 17.31 & 0.518 & 0.997 & 0.927 & 0.998 & 0.998 & 0.999 & 6.63 \\ 
		\hline\hline
	\end{tabular}}
	\caption{ Performance for detecting short epidemic segments. To conserve space, computational time~(in seconds) is reported under two settings $(L=5,\delta=2.5)$ and $(L=10,\delta=2.5)$.}
	\label{short_segment}
\end{table}

\textbf{Computational performance}: The efficiency of aPELT$_H$ does come at a cost. Table \ref{short_segment} further reports the average computational time of each method. To conserve space, we report the time under two settings $(L=5,\delta=2.5)$ and $(L=10,\delta=2.5)$, where it is seen that aPELT$_H$ in general incurs a higher computational cost compared to the testing-based methods. This is expected due to the model selection nature of aPELT$_H$. However, the increased computational cost seems rather acceptable thanks to the pruning mechanism of aPELT$_H$.
}

\section{Applications}
In this section, we demonstrate the promising performance of aPELT in two important applications, one in large-scale multiple testing and one in change-point detection of DNA copy number variation. An additional real data application in segmenting high-low volatility of oceanographic data for Canadian wave heights is given in \S5.2 of the supplementary material.

\subsection{Multiple testing with locally clustered signals}
We apply aPELT to solve multiple testing with locally clustered signals, which is an important special case of multiple testing and is studied by \cite{Zhang2011} and \cite{Cao2015}. 

Under the basic setting, a sequence of independent $p$-values $p_1,\ldots,p_n$ are observed and we need to choose between the null hypothesis $p_1,\ldots,p_n\sim U(0,1)$ and an alternative hypothesis with locally clustered signals. Specifically, the alternative hypothesis is formulated as follows: there exist change-points $0<\tau_1 <\ldots <\tau_m <n$ such that
$$p_1,\ldots,p_{\tau_1}\sim U(0,1), p_{\tau_1+1},\ldots,p_{\tau_2} \not\sim U(0,1), p_{\tau_2+1},\ldots, p_{\tau_3}\sim U(0,1),p_{\tau_3+1},\ldots,p_{\tau_4}\not\sim U(0,1),\ldots$$
or vice versa. Note that the alternative hypothesis shares a similar alternating structure as the multiple epidemic change-point problem, where the behavior of $p$-values alternates between a known common normal state~(i.e. $U(0,1)$) and epidemic states, and thus can be solved by aPELT$(\theta^o)$.

To operationalize aPELT$(\theta^o)$ for multiple testing with locally clustered signals, we model the $p$-values $p_t$ via the Beta distribution Beta$(\alpha,\beta)$. Specifically, the $p$-values on the epidemic state are modeled by a Beta distribution with parameters $\theta=(\alpha, \beta)\not=(1,1)$ and the normal state $p$-values follow Beta$(1,1)=U(0,1)$. In other words, aPELT$(\theta^o)$ is employed with $\theta^o=(1,1)$ and  the loss function $\mC(p_{(\tau_{i-1}+1):\tau_i}|\theta^i)$ is set to be twice the negative log-likelihood of $p_{(\tau_{i-1}+1):\tau_i}$ based on the Beta distribution.  After applying aPELT$(\theta^o)$ to $p_{1:n}$, we then reject the hypotheses for all the cases that are classified as epidemic states by aPELT. Note that no post-processing is needed since aPELT simultaneously estimates both the change-points and the alternating states of the sequence of $p$-values. We remark this is a general approach and can be directly applied to solve multiple testing with locally clustered signals regardless of the underlying true distribution of $p_t$.

We borrow the simulation setting from \cite{Cao2015} where $p_{1:n}$ is a sequence of $p$-values generated by two-sided tests for mean. Specifically, we have $p_t=2(1-\Phi(|y_t|))$, where $y_{1:n}$ is a sequence of independent Gaussian random variables with variance $\sigma=1$ and mean $\mu_{t}$ exhibited in Table \ref{caowu_mean}, and $\Phi(\cdot)$ is the cdf of standard normal distribution. In other words, each $p_t$ is the $p$-value of the two-sided test for $\mathbb{E}(y_t)=0$.

\begin{table}[!htbp]
	\centering
	\begin{tabular}{c|cc|c|c|c|c|c}
		\hline\hline
		Segment (\% among $n$) & 2.5 & 2.5 & 30 & 2.5 & 30 & 2.5 & 30 \\ 
		Signal strength (mean level $\mu_{t}$) & 1 & -1.5 & 0 & 1.5 & 0 & -1.5 and 1 alternating & 0\\
		State (normal/epidemic) &\multicolumn{2}{c|}{E}  & N & E & N & E & N\\
		\hline\hline
	\end{tabular}
	\caption{Mean structure with locally clustered signals that alternate between epidemic~(E) and normal~(N) states.}
	\label{caowu_mean}
\end{table}

The behavior of $p$-values $p_{1:n}$ alternates between the normal state $U(0,1)$ and different epidemic states at $\bftau^o=n(0.05,0.35,0.375,0.675,0.7)$ depending on whether $\mu_t=0$. Note that $\mu_t$ is not a constant on some epidemic states, for example, the first epidemic state $[1, 0.05n]$ and the third epidemic state $[0.675n+1,0.7n]$. Thus, the distribution of $p_t$ is not identical on those epidemic states. Furthermore, $p_t=2(1-\Phi(|y_t|))$ on epidemic states does not exactly follow the Beta distribution. Hence aPELT($\theta^o$) encounters model misspecification under the current simulation setting. However, as is shown by the numerical experiments, with the flexible Beta distribution, aPELT still gives robust and efficient performance under such misspecification.

We set $n=1000, 2000, 5000$ and compare the performance of aPELT$(\theta^o)$ with {PELT, CBS, WBS, modSaRa, BWD and the proposed procedure in \cite{Cao2015}~(hereafter CW).} For each $n$, we repeat the simulation 1000 times. We emphasize that for all detection procedures, the observed sequence is $p$-values $p_{1:n}$ instead of $y_{1:n}$. CW detects change-points and conducts multiple testing by thresholding a sequence of local scan statistics calculated via a screening window of size $k$. As a local screening method, CW has a computational cost of $O(n)$. The window size $k$ is a tuning parameter and we try both $k=\sqrt{n}$ and $k=(\log n)^2$ as suggested by \cite{Cao2015}.

{Same as aPELT, we set the loss function of PELT using the Beta distribution, thus the only difference is that aPELT explicitly explores the alternating structure of the locally clustered signals and incorporates the known normal state $U(0,1)$. {PELT, CBS, WBS, modSaRa and BWD} require additional post-processing as none automatically estimates alternating states of the $p$-values. To fix this, based on the estimated change-points, we divide the sequence of $p$-values into odd and even numbered segments, and we reject all odd numbered segments if the total length of odd numbered segments is less than $n/2$, and vice versa. This gives a slight advantage to {PELT, CBS, WBS, modSaRa and BWD} as it uses the information that the epidemic state is shorter. }

As a multiple testing problem, the ultimate interest is the realized error rate of the performed tests instead of the accuracy of the estimated change-points. Thus, we evaluate the performance of the two algorithms by the false discovery rate~(FDR), the false non-discovery rate~(FNR) and the missed discovery rate~(MDR) as suggested by \cite{Cao2015}. Specifically, the FDR follows the standard definition, the FNR is defined as the ratio of falsely accepted hypotheses and total accepted hypotheses and the MDR is defined as the ratio of falsely accepted hypotheses and total alternative hypotheses. The FNR and MDR can be used to describe the power of a multiple testing procedure, similar to the type II error rate in a single hypothesis testing setup.

The result is summarized in Table \ref{multiple_test}. In general, aPELT delivers the best performance, where it provides significantly smaller FNR and MDR than other methods while having similar FDR. Intuitively, this notable advantage of aPELT is achieved through explicitly incorporating the alternating behavior and the known normal state of the $p$-values. This example demonstrates the important application of aPELT in multiple testing and suggests that aPELT can serve as a promising and efficient tool for solving multiple testing problems with locally clustered signals.

\begin{table}[!htbp]
	\centering
	\begin{tabular}{ccccccccc}
		\hline\hline
		& $n$ & aPELT & PELT & CBS & BWD & WBS & modSaRa & CW $k=(\log n)^2$  \\
		\hline
		FDR & 1000 & 6.73 & 3.82 & 9.58 & 15.69 & 24.08 & 8.52 & 2.35 \\ 
		& 2000 & 5.45 & 3.62 & 8.15 & 12.08 & 21.19 & 13.74 & 7.19 \\ 
		& 5000 & 2.02 & 1.90 & 3.33 & 5.63 & 4.35 & 15.29 & 3.90 \\ 
		FNR & 1000 & 3.74 & 5.87 & 5.43 & 4.50 & 6.77 & 6.16 & 8.23 \\ 
		& 2000 & 1.44 & 2.52 & 2.22 & 1.84 & 3.06 & 4.47 & 4.65 \\ 
		& 5000 & 0.39 & 0.56 & 1.01 & 0.82 & 0.37 & 2.17 & 3.73 \\ 
		MDR & 1000 & 35.23 & 56.58 & 52.12 & 42.20 & 60.83 & 59.78 & 80.95 \\ 
		& 2000 & 13.30 & 23.45 & 20.43 & 16.58 & 25.20 & 41.99 & 42.04 \\ 
		& 5000 & 3.50 & 5.13 & 9.25 & 7.28 & 3.27 & 19.88 & 31.31 \\ \hline\hline
	\end{tabular}
	\caption{Performance for multiple testing with locally clustered signals. The performance of CW $k=\sqrt{n}$ is slightly worse than CW $k=(\log n)^2$ and thus is omitted.}
	\label{multiple_test}
\end{table}

\subsection{DNA copy number variation}
In this section, we apply {aPELT$_H$, PELT, CBS, WBS, modSaRa and BWD to analyze CNVs in aCGH data for the NCI-60 cell lines reported in \cite{Varma2014}. We give the detailed result for two representative sequences, OVCAR-3~(here) and SN12C~(in the supplementary material).} The result for other sequences is qualitatively similar and thus is omitted. 

\begin{figure}[h!]
		\centering
		\includegraphics[scale=0.65, angle=270]{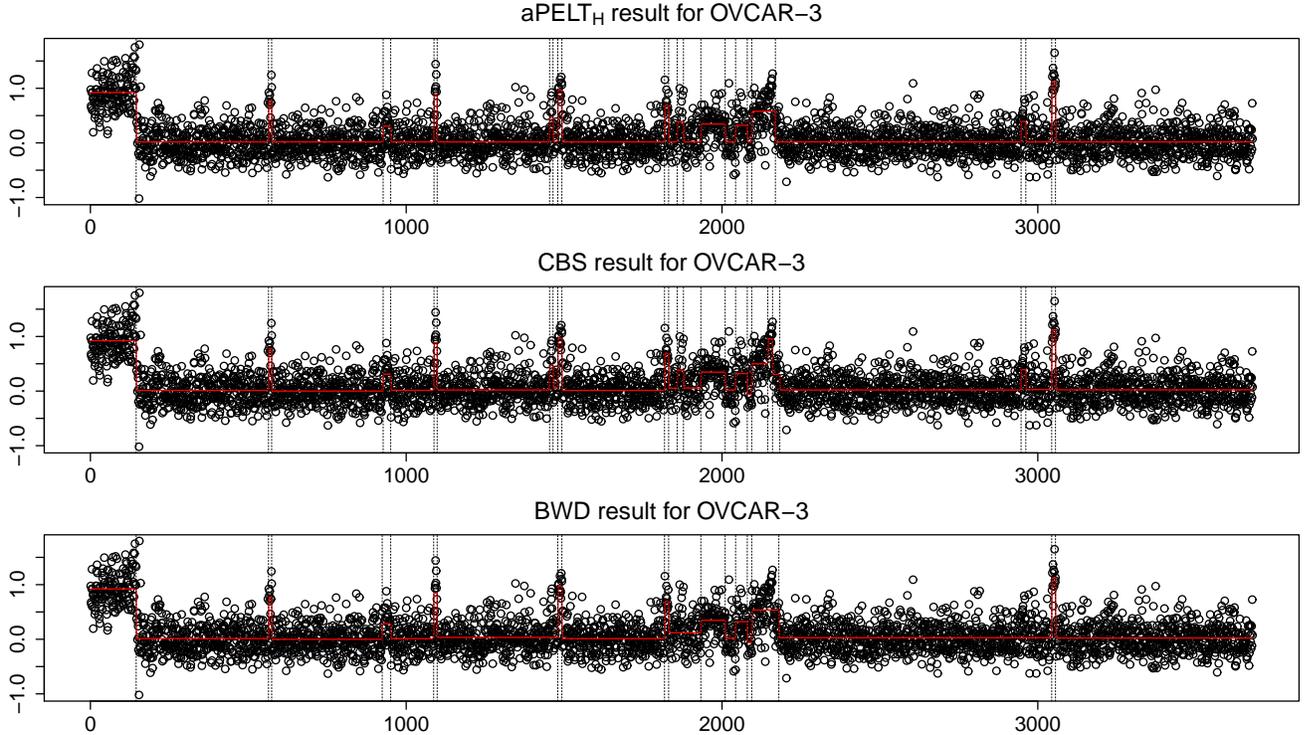}
		\caption{ Estimated change-point locations and mean vectors by CBS, BWD and aPELT$_H$.}
		\label{DNAcnv_SN12C}
\end{figure}

{The estimated change-points for OVCAR-3 are visualized in Figure \ref{DNAcnv_SN12C}. To conserve space, we only give the result for CBS, BWD and aPELT$_H$, which are the top performers in Section 4.2.} Overall, the three algorithms provide similar segmentation results. Note that the estimated mean vector $\hat{\mu}_{1:n}$ by all algorithms exhibit the alternating structure, which justifies the use of aPELT$_H$.
	
To further confirm the advantage of aPELT$_H$, we compute the BIC of the estimated segmentation for OVCAR-3 given by each method. Denote $\hat{m}$ as the number of estimated change-points. For CBS and BWD, the number of model parameters~(denoted by $p$) is $2\hat{m}+2$, accounting for $\hat{m}$ change-points and $\hat{m}+1$ normal distributions {with homoscedasticity}. For aPELT$_H$, $p=(\hat{m}+2+ \text{No. of epidemic states})$, as all normal states share the same mean parameter $\tilde{\mu}^o$ due to the imposed epidemic constraint of the mean vector. {The BIC is $1065.6$ by CBS, $1076.6$ by BWD and $972.7$ by aPELT$_H$.} Thus, aPELT$_H$ offers the best change-point model~(based on BIC) by explicitly incorporating the epidemic mean structure. The same result also holds for OVCAR-3.

For the optimization of aPELT$_H$, Figure S.5 of the supplementary material plots the estimated normal state parameter $\tilde{\mu}^o$ by aPELT$_H$ along with the function $F^*(n;\mu^o)$ for OVCAR-3 and SN12C. The result confirms that $F^*(n;\mu^o)$ is piecewise smooth and that aPELT$_H$ successfully achieves the minimum value of $F^*(n;\mu^o)$, with the normal state estimates $\tilde{\mu}^o=0.018$ for OVCAR-3 and $\tilde{\mu}^o=-0.093$ for SN12C, both close to 0.

\vspace{-0.2cm}
\vspace{-0.6cm}

\section{Discussion}
In this paper, we generalize the classical (single) epidemic change-point detection problem to a more realistic multiple epidemic change-point setting. The interest is to estimate the unknown number and locations of change-points and the alternating states of the observed sequence.

To explicitly incorporate the alternating structure of the new problem, we propose a novel model selection based approach for \textit{simultaneous} inference on both change-points and alternating states. A two-stage alternating pruned dynamic programming algorithm~(aPELT) is further developed, which conducts efficient and accurate optimization of the model selection criterion. The favorable performance of aPELT is demonstrated via extensive numerical experiments and meaningful applications to multiple testing, DNA copy number variation and oceanographic study.

{Due to its model selection nature, aPELT can be naturally extended to epidemic change-point detection in multivariate time series. Specifically, we can modify the model selection criterion to reflect the multivariate nature of the time series, which can be done by suitably adjusting the measure of model fit $\mathcal{C}$~(e.g.\ to the log-likelihood of multivariate Gaussian distribution) and the penalty terms $P^o, P^1$. The optimization can be carried out in exactly the same way as that for the univariate time series. See Section \S2 of the supplementary material for a more detailed discussion.}

Another conceptually feasible and {promising} solution to the proposed multiple epidemic change-point problem is a hidden Markov model~(HMM) with a structured transition matrix, which dictates the alternating transition between the normal state and epidemic states. {We refer to \cite{Wang2014} and \cite{Steibel2015} for some recent works on HMM. A rigorous and detailed study of this direction is beyond the scope of this paper and we leave it for future research. }

\section*{Acknowledgements}
The authors thank the associate editor and two anonymous referees for their comments that helped improve the quality and presentation of the paper. Zhao's research is supported in part by National Science Foundation grant DMS-2014053. Yau's research is supported in part by
grants from HKSAR-RGC-GRF 14302719 and 14305517.

\setlength{\bibsep}{5pt plus 0.3ex}
\bibliographystyle{apalike}
\bibliography{CP_Reference}
\end{document}